\input{aipcheck}

\documentclass[
    ,final            % use final for the camera ready runs
  ]
  {aipproc}

\layoutstyle{6x9}
\usepackage{amsmath}
\usepackage{epsfig}
\def\nostrocostrutto#1\over#2{\mathrel{\mathop{\kern 0pt \rlap
  {\raise.2ex\hbox{$#1$}}}
  \lower.9ex\hbox{\kern-.190em $#2$}}}
\def\lsim{\nostrocostrutto < \over \sim}   %less or around ...  
\def\gsim{\nostrocostrutto > \over \sim}   %greater or around...

\begin{document}

\title{Scalar Mesons in B-decays}


\author{Peter Minkowski}{
 address={Institute for Theoretical Physics, University of Bern, CH-3012 Bern,
Switzerland}
}

\author{Wolfgang Ochs}{
  address={Max-Planck-Institut f\"ur Physik (Werner-Heisenberg-Institut),\\ 
F\"ohringer Ring 6, D-80805 M\"unchen, Germany}
}

\begin{abstract}
 We summarize some persistent problems in scalar spectroscopy and discuss what
could be learned here from charmless B-decays. Recent experimental
results are discussed in comparison with theoretical
expectations: a simple model based on penguin dominance leads to various
symmetry relations in good agreement with recent data; 
a factorisation approach yields absolute predictions of
rates. For more details, see \cite{talk}.
\end{abstract}

\maketitle

%%%%%%%%%%%%%%%%%%%%%%%%%%%%%%%%%%%%%%%%%%%%
%% MAINMATTER
%%%%%%%%%%%%%%%%%%%%%%%%%%%%%%%%%%%%%%%%%%%%

\section{WHY STUDYING SCALARS IN B-DECAYS}
There are various reasons for studying scalar particles in B-decays:

\emph{1. Dominance of $S$-wave resonances with little background from
crossed channels}\\
In $(B\to 1,2,3)$-decays the masses of (1,2) resonances 
can extend to $M\lsim 5$ GeV.
Then there is little overlap with resonances in crossed channels (2,3) or
(1,3). This is very different from $D$ decays where 
resonance masses extend only up to $\sim1.5$ GeV and in general there is
a large overlap.
Furthermore, in the final 2-body systems $S$-wave interactions are dominant.

\emph{2. New source of glueballs}

The elementary subprocess $b\to sg$ with an isolated gluon is rather well
understood theoretically and 
is described by a penguin diagram. The decay rate has been calculated in
next-to-leading order of perturbative QCD as \cite{greub}
\begin{equation}
{ B} (b\to sg) = (5\pm 1)\times 10^{-3}. 
\end{equation}
The gluon may give rise to production of a glueball which could show up as
a resonance in the system $X$ of 2-body decays  $B\to K^{(*)}+X$. 
This process adds to the other well known
gluon rich processes like: central production in $pp$ collisions, $J/\psi\to
\gamma X$ and $p\bar p$ annihilation near threshold.  

\emph{3. Non-charm final states with strangeness}\\
The decays $b\to sq\bar q$ are dominated again by the gluonic
penguin process whereas the electroweak tree diagrams occur at the level of
20\% only. In the leading penguin approximation the decays $b\to s u\bar u,
\ s d\bar
d ,\ s s\bar s $ occur with the same fraction and have been calculated to amount 
to $\sim 2\times 10^{-3}$ each. In the corresponding hadronic 
2-body final states $B\to xy$, if $x$ and $y$ are members of $SU(3)$
multiplets $X,Y$ each, one obtains various symmetry relations \cite{mo1}. Hopefully,
this will ultimately identify the members of the lightest scalar nonet and
the mixing properties.     

\section{PROBLEMS OF LIGHT SCALAR MESON SPECTROSCOPY}
The interest in light scalar mesons originates from the following
expectations:\\
\emph{1. The existence of glueballs}\\
This is a requirement from the first days of QCD 
and may be the most urgent open problem of the theory at the fundamental
level. In lattice QCD, quenched approximation, 
the lightest glueball appears in the $0^{++}$ channel
with a mass of 1400-1800 MeV \cite{bali}. The effect of unquenching is under
study but realistic estimates are still difficult, especially 
because of the large quark masses. An alternative QCD approach is based on
QCD sum rules \cite{narison} where the lightest glueball is centered
around 1000-1400 MeV.  

\emph{2. Multiplets of $q\bar q$ and exotic bound states}\\
There is no general consensus on the members of 
the lightest $q\bar q$ nonet, i.e. the
parity partner of $\pi,K,\eta,\eta'$. In addition, there is the possibility
of tetraquarks  \cite{jaffe}, bound states of di-quarks.
%which explain various common features of 
%meson and baryon spectroscopy \cite{wilcek}.
%In order to resolve these puzzles, it is important to determine the
%couplings, including the sign, into the various decay channels. 

The list of scalar particles provided by the PDG \cite{pdg} 
with mass $M\lsim 1.8$ GeV 
includes \\
$I=0$: $f_0(600)$ ({\rm or} $\sigma$),  $f_0(980)$,
$f_0(1370)$,  $f_0(1500)$,  $f_0(1710)$;\\ 
$I=\frac{1}{2}$: $\kappa(900)$ (?),  $K^*_0(1430)$; \\
${ I=1}$:  $a_0(980)$,  $a_0(1450)$.

There are two typical scenarios for the classification of these states:

\emph{I. One nonet below and one above 1 GeV}\\
The nonet of lower mass includes $\sigma,\ \kappa,\ f_0(980),\ a_0(980)$,
either $q \bar q$ (see, for example, Ref. \cite{morgan} 
and Van Beveren \cite{conf}) or $qq\bar q\bar q$ 
\cite{jaffe} bound states. The higher mass states could then make a $q\bar
q$ nonet with members $K^*_0(1430)$ and $a_0(1450)$; in the isoscalar sector
the three states $f_0(1370)$,  $f_0(1500)$ and $f_0(1710)$
could be, as originally proposed in \cite{amsler}, a superposition of 
the glueball and the two members of the isoscalar nonet.

\emph{II. One nonet above 1 GeV}\\
In this scheme the $\sigma$ and $\kappa$ with the parameters given 
are not considered as physical states to be classified along the lines we
discuss here. The $q\bar q$ nonet is rather formed by 
 $a_0(980)$ (or also $a_0(1450)$), $ f_0(980)$,  $K^*_0(1430)$
and $f_0(1500)$ \cite{klempt, mo2} whereas two higher mass nonets
including $f_0(1370)$ have been proposed in \cite{anisovich}.
The $\pi\pi$ $S$-wave is interpreted as being dominated by a very broad
object, centered around 1 GeV, the lower part could be responsible for the
$\sigma(500)$ effect. This broad state ($\Gamma>500$ MeV) has been 
proposed as representing the isoscalar glueball 
by various arguments \cite{anisovich, mo2}.   

There are states whose identity is in doubt as can be seen by the
large uncertainty in mass and width estimated by the PDG:
 $\sigma(500)$, % ($M=400-1200$ MeV, $\Gamma=600-1000$ MeV); 
$f_0(1370)$
%($M=1200-1500$ MeV, $\Gamma=200-500$ MeV) 
with no single branching ratio or
ratio of such numbers accepted by PDG and finally $\kappa$ or $K^*(800)$
%$M=700-900$ MeV, $\Gamma=400-800$ MeV, 
not carried in the main listing of
PDG. We will add a few remarks on these problematic states
which will be of relevance for our discussion of $B$ decays.

\subsection{Isoscalar channel: $f_0(600)$ or $\sigma$ and $f_0(1370)$}
Most definitive experimental results on these states can be obtained 
from the $2\to2$  scattering processes 
$\pi\pi\to\pi\pi,K\bar K,\eta\eta$ applying 
an energy independent partial wave
analysis (EIPWA); in this case unitarity provides important constraints in the
full energy range. Recently, results on $D$ and $B$ decays as well as $p\bar
p\to 3$ particles with higher statistics became available.
There is no general constraint on the mass dependence of the amplitude
which can be affected by various dynamical effects.
So far, in these processes no EIPWA over the full energy range has been
performed, so an optimal description of data for  a particular model
parametrization is selected. A promising 
new approach towards EIPWA in $D$-decays has
been presented at this conference by Meadows \cite{meadows}. 

Concerning the $\pi\pi$ interaction there is a general consensus that there
exists indeed a broad state with the width of the order of the mass, but the
parameters depend on the mass range considered, 
a feature which is known already since about 30 years.

\emph{1. Low mass range $M_{\pi\pi}\lsim 0.9\ldots 1.2$ GeV.}\\
In this region the complex $\pi\pi$ amplitude moves along the unitarity circle
to its top (phase $90^\circ$) where a rapid circular motion follows from
$f_0(980)$.
An early analysis has been performed by the Berkeley
collaboration \cite{protopopescu}, they found a state, $\sigma$, with
$M_\sigma=660\pm100$ MeV, $\Gamma_\sigma=640\pm140$ MeV. Recently, results
from $D$-decays by E791 \cite{e791}, FOCUS \cite{focus} and from 
$J/\psi\to \omega\pi\pi$ by BES \cite{bes} have been interpreted in terms of
a $\sigma$ with similar mass, although good fits based on a $K$ matrix 
parametrization have
been obtained without such a state \cite{focus}. On the theoretical side,
parametrizations of such data using the low mass $\chi PT$ constraints
lead to a low mass pole with  $M_\sigma \sim 450$ MeV and 
$\Gamma_\sigma \sim 450\ldots 600$ MeV (see, e.g. Refs.
\cite{oop,cgl} and the reports by Bugg and Pelaes 
\cite{conf1}). 

\emph{2. Extended mass range $500 \leq M \leq 1800$ MeV}\\
In case of a broad state the parameters should be determined from the energy
interval where its influence is important and this includes the inelastic
region above 1 GeV.

All analyses of  $\pi\pi$ scattering in this region find again one broad state, 
but with a higher mass than
before, in a range around 1000 MeV and with large width $>500$ MeV.
The first analysis along these lines goes back again 30 years \cite{hyams} 
and in Table 1
we list the pole positions from K matrix fits of various analyses.
The fits by Estabrooks \cite{estabrooks} 
refer to the four solutions of an EIPWA of elastic
$\pi\pi$ scattering \cite{em} as well as of the $\pi\pi\to K\bar K$ reaction.
In all solutions of the  EIPWA the 
$S$-wave amplitude above 1 GeV follows a circular
path with some inelasticity in the Argand
diagram ( $Im\ T$ vs. $Re\ T$) which can be fitted by a broad resonance.
Superimposed is a smaller circle corresponding to a resonance \cite{estabrooks}
 with parameters close to what is known today as
$f_0(1500)$.  
No additional pole, such as
$f_0(1370)$, is seen in this analysis. A similar picture is found
\cite{mo2} for the
inelastic channels $\pi\pi\to\eta\eta$ and $\pi\pi\to K\bar K$ 
comprising the broad background and $f_0(1500)$  with the interference pattern 
\begin{equation}
\begin{tabular}{lll}
$\pi\pi\to K\bar K$: & background - $f_0(1500)$ & {constructive interference} \\
$\pi\pi\to\eta\eta: $& background - $f_0(1500)$ & {destructive interference} \\
\end{tabular}
\end{equation}
 This broad state is seen in a variety of processes
and has been dubbed $f_0(1000)$ in \cite{amp}. 
Later arguments have been presented that this broad state be a glueball
\cite{anisovich, mo2}. 
This state also appears in decay processes although it may happen that
the higher mass tails are suppressed for dynamical reasons. As an example, we
quote 
the study by BES \cite{beskk} of the 
final state $J/\psi\to \omega K^+K^-$ where the large $S$-wave background
(``$\sigma$'') extends up to about 2 GeV.
A significant flat
background has also been observed recently in the gluon rich channel 
$J/\psi\to \gamma K\bar K$ by BES \cite{besgammakk}. 

Apparently, the state $f_0(1370)$ shows up if 
reactions other than (1)-(3) in Table 1 
without unitarity constraints are included in the
fits. Whereas $f_0(980)$, $``f_0(1000)$'' and $ f_0(1500)$ are clearly seen
as circles in the Argand diagrams, no such circle has ever been shown 
to exist for
$f_0(1370)$. Before such a behaviour is demonstrated, this state could
hardly be considered as
established. The strong interference between background and $f_0(1500)$,
leads to very different mass spectra, depending on the relative phase, 
which could easily simulate a ``new state'' $f_0(1370)$. 
\begin{table}[ht]
\begin{tabular}{llll}
\hline
Authors                &   mass (MeV) &    width (MeV)&    channels \\
\hline
CERN-Munich \cite{hyams} & 1049      &  500   & 1\\
Estabrooks \cite{estabrooks} & 750  &   800-1000 & 1,2 \\
Au, Morgan \& Pennington \cite{amp} & 910 & 700    & 1,2,5\\
Anisovich and Sarantsev \cite{anisovich}& 1530 & 1120&  1,2,3,4\\
%${\bf I=0}$ &$f_0(600)$ ({\rm or} $\sigma$) & $f_0(980)$ &
%$f_0(1370)$ & $f_0(1500)$ &  $f_0(1710)$ &  $f_0(2020) ?$\\
\hline
\end{tabular}
\caption{Position of broad state in the $T$ matrix of $\pi\pi$ scattering 
according to various K matrix fits to data from reactions (1)  $\pi\pi\to \pi\pi$,
 (2)  $\pi\pi \to K \bar K$, (3) $\pi\pi\to \eta\eta, \eta\eta'$, (4) 
$p\bar p$ annihilation and (5) $J/\psi$ decays }
\label{tab:pole}
\end{table}

We conclude that there is indeed a broad state in
the isoscalar channel with decays into various 2-body final states
but there is no standard form for its line shape.
Different results on its mass emerge depending on whether the analytic
parametrization is fitted to a small or a large mass range (corresponding to
either a half resonance circle or an almost  full circle) leading either to
$\sigma(500)$ or ``$f_0(1000)$''.
There can be little doubt that both
results refer to the same state. Studies along path 1 should ultimately 
extend their parametrization to include higher
mass inelastic channels, especially the EIPWA results by Estabrooks, whereas
the analyses along path 2 should include the very low mass $\pi\pi$ data
as well.

\subsection{Isospin $I=\frac{1}{2}$ channel: $\kappa(800)$ and $K^*_0(1430)$}
The elastic $K\pi$ scattering up to 1700 MeV 
has been studied some time ago
by an experiment at SLAC \cite{slac} and the LASS experiment \cite{lass}.
The $S$ wave phase shifts have been parametrized in terms of $K^*_0(1430)$ 
with a small inelasticity $<10\%$ starting only above the inelastic threshold
$M_{K\pi}\gsim M_{K\eta'}\sim 1450$ MeV and a slowly varying background
with an effective range formula. This background 
phase in the considered range does not exceed about 50$^\circ$ and insofar
it is a phenomenon quite different from the background in 
$\pi\pi$ scattering where the
background phase reaches 90$^\circ$ below the first scalar resonance $f_0(980)$.
We do not want to enter here into the discussion about a possible
state $\kappa$ 
but point to
different characteristics of the $K\pi$ amplitude in 
elastic scattering and decay relevant to our later discussion.
For a theoretical analysis, see B\"uttiker et al. \cite{buttiker}. 

\begin{figure}[h!]
%\begin{center}
%\vspace{-3.0cm}
%\includegraphics[angle=+90,width=15cm,bbllx=240,bblly=88,bburx=427,bbury=749]{p-001074-Model.EPS}
%\includegraphics[angle=-90,width=14cm]{NF-dla.ps}
\includegraphics[width=5cm]{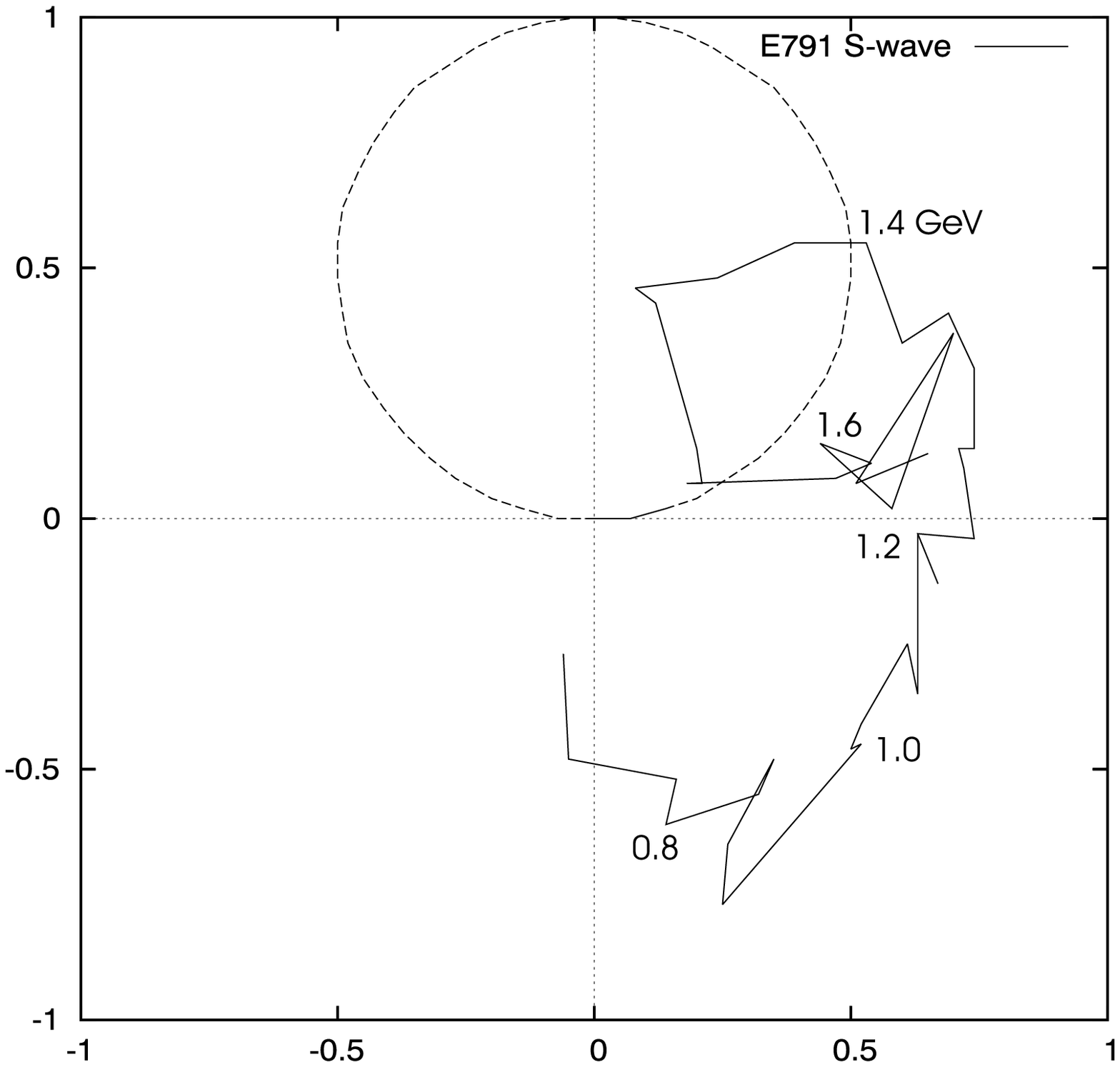}
%\end{center}  
%
%\vspace{-5cm}
\caption{
Energy independent partial wave analysis of the $I=\frac{1}{2}$ $K\pi$
$S$-wave using E791 data ``$c_i$'' and form 
factors $F_{Di}$ \protect\cite{meadows}.
 Plotted are the rescaled quantities $S_i=c_iF_{Di}\frac{q}{\protect\sqrt{s}}$
(arbitrary units) to be compared with the elastic unitarity circle
for $K\pi\to K\pi$ scattering.
}
\label{fig:eipwae791}
\end{figure}

In weak decays like $D\to K\pi\mu\nu$ the $K\pi$ phase equals the one in
elastic scattering according to the Watson theorem and this is nicely born
out by the data (FOCUS \cite{goebel}). If rescattering effects are small,
then the Watson theorem is still applicable in purely hadronic decays and a
recent example for this behaviour is $B\to J/\psi K\pi$ measured by BaBar
\cite{babarkpi}. On the other hand, in $D\to K\pi\pi$ the $K\pi$ phases
determined by E791
\cite{meadows} follow the trend as in elastic scattering below the inelastic
$K\eta'$ threshold $M\sim1400$ MeV, but with a relative shift of about
70$^\circ$.
The Argand diagram in Fig. \ref{fig:eipwae791})
shows that the resonance circle 
related to $K^*_0(1430)$ is much smaller than the circle related to the
background, which contrasts to elastic scattering with circles of comparable
radii.
 Therefore the LASS parametrization does not represent
the decay amplitude in an energy region beyond 1400 MeV.  
% 
%\begin{figure}
%  \includegraphics[height=.3\textheight]{wave}  
%  \caption{The four solutions for the isoscalar $\pi\pi$ S-wave above 1 GeV}
%%  \caption{Picture to fixed height}
%\end{figure}
%
%\begin{figure} [htb]
%\begin{center}
%\centerline{\hspace{1.2cm}
%\epsfig{file=wave.eps,width=5.1cm}
%\vspace{-0.1cm}
%\caption[]{ The four solutions for the isoscalar $\pi\pi$ S-wave above 1
%GeV}
%\end{center}
%\end{figure}

%\vspace{-5cm}

\section{$B$-decays: experimental results on scalars}
The $B$ branching ratios for the
following scalar particles have been measured,
for later comparison we present the results corrected for unseen channels,
all in units of $10^{-6}$.

\emph{Isospin I=1: $a_0(980)$}\\
So far only upper limits have been reported by BaBar
\cite{babara0},\footnote{After the conference results by Belle \cite{bellea0}
became available which confirm the tight upper bounds for 
$a_0$ production: $B(B\to a_0^-K^+) < 1.6\times 10^{-6}$ for the $\eta\pi$ channel. } 
see Tab. \ref{babara0dat}.
%Using
%$B(a_0\to\eta\pi)=0.85$ one finds
\begin{table}[htb]
\begin{tabular}{lllllc}
\hline
$B^0\to K^0a^0_0$  & <\ 9.2 & \qquad\qquad & $B^+\to K^+a^0_0$ & <\ 2.9
\\%& all numbers\qquad\qquad\\
$B^0\to K^+a^-_0$  & <\  2.4 & \qquad\qquad &  $B^-\to \bar K^0a^-_0$ & <\ 4.6
\\%&$ \times 10^{-6}$ \qquad\qquad\\
\hline
\end{tabular}
\caption{$B$ decays into $a_0(980)$ \cite{babara0} corrected using
$B(a_0\to\eta\pi)=0.85$; all numbers $\times 10^{-6}$.}
\label{babara0dat}
\end{table}

\emph{Isospin $I=\frac{1}{2}: K^{*}_0(1430)$}\\
Total decay rates derived from BaBar \cite{babarkst} and Belle data
\cite{bellekst} are given in 
Tab. \ref{tab:kst}.
% corrected for unseen decay modes (using $B(K^*_0\to K\pi)=0.93$) 
\begin{table}
\begin{tabular}{lcc}
\hline
                         &  BaBar \protect\cite{babarkst}      
& \qquad Belle \protect\cite{bellekst}  \\
\hline
$B^0\to K^{*0}_0 \pi^0$ \qquad\qquad  & $12.7\pm5.0$ &\qquad  $9.8\pm 2.7 $\\
$B^0\to K^{*+}_0 \pi^-$ \qquad\qquad & $36.1\pm12.2$ &\qquad $16.4 \pm 5.3 $\\
$B^+\to K^{*0}_0 \pi^+$ \qquad\qquad & $37.0\pm4.4$ &% 
\qquad (I) $ 45 ^{+15.3}_{-11.1}$ \qquad (II) $ 8.3^{+3.9}_{-2.6}$\\
\hline
\end{tabular}
\caption{$B$ decays into $K^*_0(1430)$ corrected for unseen modes using 
$B(K^*_0\to K\pi)=0.93$; units in $10^{-6}$}
\label{tab:kst}
\end{table}
Belle, in a full Dalitz plot analysis using an isobar model ansatz 
finds two quite different solutions 
in $B^+\to K^{*0}_0\pi^+$ corresponding to different interferences with a 
coherent background amplitude. Babar is inserting the LASS parametrization
for the $K\pi$ phase in a larger energy interval up to $M\sim M_D$
and is then left with only one solution.
As discussed above for $D\to K\pi\pi$ the behaviour of the $K\pi$ amplitude
above 1400 MeV could be quite different from elastic scattering and a more
general ansatz in this mass region seems appropriate. 
The situation is quite analogous to $f_0(980)$ in $\pi\pi$ interactions
where the interference pattern of $f_0$ and background changes from
one reaction to another. It will be therefore important to clarify the 
existence of two
solutions and to possibly exclude one of them by physical arguments.

\emph{Isospin I=0: $f_0(1500)$}\\
Both Belle \cite{bellekst} and BaBar \cite{babarkk}
see a peak in the $K^+K^-$ Mass spectrum in $B\to
(K^+K^-)K$. The mass and width are consistent with
$f_0(1500)$. However, no signal in the corresponding $\pi\pi$ decay channel
is observed despite the ratio of branching ratios $\Gamma(f_0\to K\bar K)/
\Gamma(f_0\to \pi\pi)= 0.241\pm0.028$ \cite{pdg} 
is favourable for the $\pi\pi$ channel.
Therefore both collaborations suggested the existence of a new state,
$f_X(1500)$ or $X(1500)$. 

\begin{figure}[t!]%[hbt]
\vspace{-1.0cm}
%
%\begin{center}
\begin{minipage}[t]{10cm}
\centering\includegraphics[angle=0,width=10cm,bbllx=50,bblly=60,bburx=700,bbury=532]{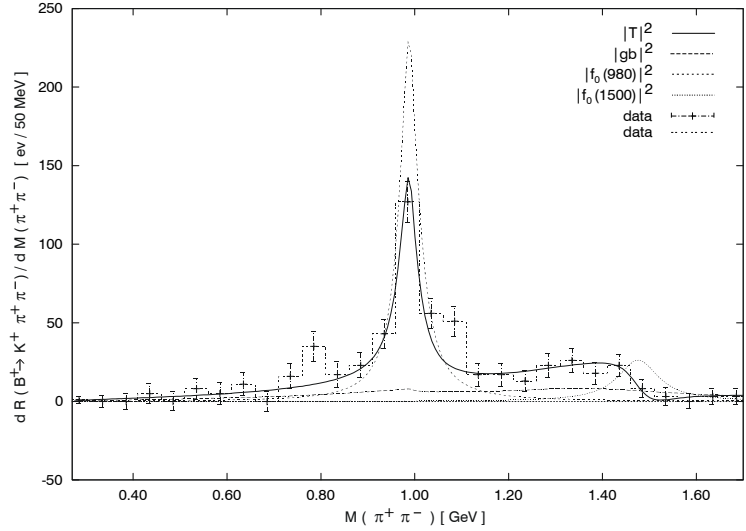}\\ %  {pipi+64.ps}
%\mbox{\includegraphics[angle=0,width=10cm,bbllx=50,bblly=60,bburx=700,bbury=532]{pipi+64bw.ps}}\\ %  {pipi+64.ps}
%
\end{minipage}
%
%\begin{minipage}[t]{10cm}
%\mbox{\includegraphics[angle=0,width=10cm,bbllx=50,bblly=60,bburx=700,bbury=532]{KKbar+67bw.ps}} %{KKbar+67.ps}
%\centering\includegraphics[angle=0,width=10cm,bbllx=50,bblly=60,bburx=700,bbury=532]{KKbar+67bw.ps} %{KKbar+67.ps}
%
%\end{minipage}
%\end{center}  
%
\vspace{-1.0cm}
\caption{$\pi^+\pi^-$ mass spectrum in $B$-decays
 (BELLE
\protect\cite{bellekst}) in comparison with a model 
including $f_0(980),\ f_0(1500)$ and a broad glueball
($gb$).
Also shown are the individual resonance terms.
 The background ($gb$) in this fit 
interferes destructively with both $f_0(980)$
and $f_0(1500)$. 
%according to the model amplitudes
%(Eqs. (\protect\ref{massspectrum}) and (\protect\ref{background})).
}
%end fig. a
\label{fig:pipm}
\end{figure}

\begin{figure}[h!]%[hbt]
%\begin{center}
%\vspace{-1.0cm}
%\includegraphics[angle=0,width=10cm]{KKbar+67bw.ps} %{KKbar+67.ps}
%\hspace*{-1.275cm}
%\begin{minipage}[t]{10cm}
\centering\includegraphics[angle=0,width=10cm,bbllx=50,bblly=60,bburx=700,bbury=532]{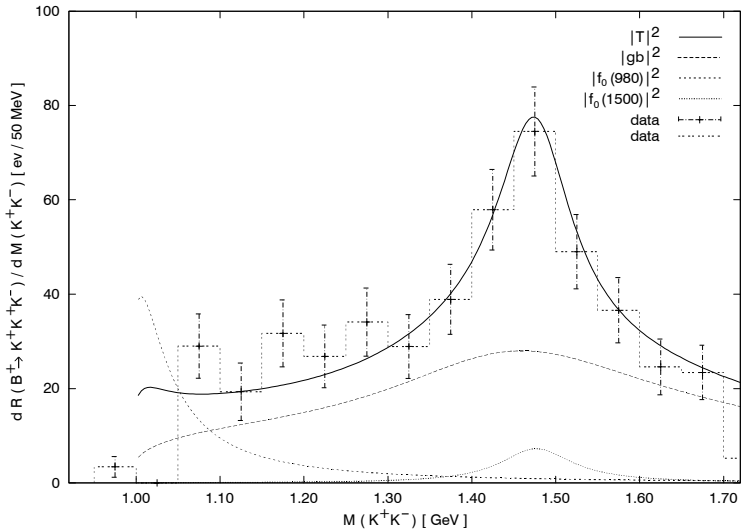} %{KKbar+67.ps}
%\end{minipage}
%
%\end{center}  
\vspace{-2.5cm}
\caption{$K^+K^-$ mass spectrum in $B$-decays
(BELLE \protect\cite{bellekst}) 
in comparison with the model amplitude, see Fig.
(\protect\ref{fig:pipm}). Here $f_0(1500)$ interferes constructively  
with the background.
}
\label{fig:kpm}
\end{figure}

In a previous work \cite{mo1} studying the Belle data \cite{bellekst} 
we argued that  these phenomena 
are naturally explained by the existence of a broad
background which interferes with $f_0(1500)$: 
constructively in $K^+K^-$ giving rise to the observed peak
but destructively in $\pi^+\pi^-$ leading to a vanishing signal.
In our
analysis we have represented the mass spectra as a superposition of three
components $f_0(980),\ f_0(1500)$ and a broad resonance as background,
which fits the data well, see Figs. % $bg$
%\begin{align}
%\frac{d\Gamma}{dm}= &qp|c_1|^2|T_{bg}+c_2T_{f_0}S_{bg}+c_3T_{f_0'}D_{bg}|^2
%\label{massspectrum}\\
%    S_{bg}=& e^{2i\delta_{bg}};\qquad T_{bg}=|T_{bg}|e^{i\delta_{bg}}
%\label{background}
%\end{align} 
%where $T$ denotes an inelastic Breit Wigner resonance. This superposition
%of amplitudes corresponds to the multiplication of $S$-matrices 
%(``Dalitz and Tuan'') and is compared with data in Fig.'s
\ref{fig:pipm},\ref{fig:kpm}. 
%The interference in $K^+K^-$ is constructive and in
%$\pi^+\pi^-$ destructive. 
This interference pattern is the same as in inelastic
$\pi\pi\to K\bar K, \eta\eta $ \cite{mo2} and elastic $\pi\pi$ scattering
\cite{estabrooks} (see above).   

These signs are consistent with our hypothesis \cite{mo2} that the
background represents a broad glueball (flavour singlet) 
 with mass in the 1 GeV region or above
interfering with $f_0(1500)$, which is close to a flavour octet state 
%We assume the
%mixing as in the pseudoscalar sector with correspondence
%$\eta'\Leftrightarrow f_0(980)$ and $\eta \Leftrightarrow f_0(1500)$
according to the considerations in \cite{klempt,mo2}. 

Both collaborations find two solutions for the $B\to K f_0(1500)$ rate 
corresponding to different
interference signs with the background. 
From the total charmless and the partial
fractions we obtain the branching ratios in Tab. \ref{bf01500}.
According to our model Sol. II is the physical solution.
%using $B(f_0(1500)\to K^+K^-)=0.043$:
\begin{table}[thb]
\begin{tabular}{lll}
\hline
$B(B\to f_0(1500)K)$               \qquad          &  Belle ($B^+$) \qquad
               & \qquad  BaBar $(B^0)$  \\
\hline
Sol. I\quad  ($bg-f_0)$ \qquad\qquad  & $471.8\pm51.3$ &\qquad  $223\pm 42$ \\ 
Sol. II ($bg+f_0)$ \qquad\qquad & $61.1\pm 14.4$ &\qquad $29.9\pm1 3.7 $ \\
\hline
\end{tabular}
\caption{$B$-decay rates ($\times 10^{-6}$) into $f_0(1500)K$ 
(total rates using $B(f_0(1500)\to
K^+K^-)=0.043$).
%\vspace{-0.7cm}
} 
\label{bf01500}
\end{table}

\emph{Isospin I=0: $f_0(980)$}\\
$f_0(980)$ was the first scalar particle observed in $B$-decays
and the results obtained by
the heavy flavor averaging group (HFAG) \cite{hfag} 
are presented in Tab. \ref{f0rates}.
In these decay channels there is again some background contribution as in case of
$f_0(1500)$, so we expect two possible solutions corresponding to
different interference signs. Note the negative interference in our fits in
Figs. \ref{fig:pipm},\ref{fig:kpm}, 
also observed in $J/\psi\to \phi\pi\pi$ by DM2
\cite{dm2}. 
It would be important to study the possibility of a second solution 
besides the one in Tab. \ref{f0rates}.
%The results obtained by
%the heavy flavor averaging group (HFAG) \cite{hfag} for the  decay 
%channel $f_0(980)\to \pi\pi$ and for
% the total rate using $\Gamma(f_0(980)\to \pi\pi)/\Gamma_{tot}\sim 0.8$
%are 

\emph{Other results on scalars}\\
$f_0(1370)$:
A $2\sigma$ signal has been observed by Belle \cite{bellekst}, not so by
BaBar \cite{babarkst}.

$\sigma(600)$:
No obvious peak near threshold is visible as in $D\to 3\pi$.
For our discussion of
scalars it would be interesting to 
obtain  the rate for $B\to \sigma K$ (or a limit).

$\kappa(800):$
A $K\pi$ background in $K\pi\pi$ has been observed by Belle \cite{bellekst}
but a fit with a $\kappa$ particle was not successful.
 
$a_0(1450)$:
This state has not been seen yet.

\begin{table}[h!]
\begin{tabular}{ccc}
\hline
                         &  $f_0(980)\to \pi\pi$ \qquad  &  $f_0(980)\to all$ \\
\hline
$B(B^+\to f_0(980) K^+)$ \qquad\qquad  & $13.2\pm 1.6$   & 16.5  \\
$B(B^0\to f_0(980) K^0)$ \qquad\qquad & $ 8.25 \pm 1.5 $ & 10.3 \\
\hline
\end{tabular}
\caption{$B$-decay rates ($\times 10^{-6}$) into $f_0(980)K$,  for 
$f_0(980)\to \pi\pi$ and total rate using  $\Gamma(f_0(980)\to
\pi\pi)/\Gamma_{tot}\sim 0.8$.
} 
\label{f0rates}
\end{table}
\section{$B$-decays into scalars: theoretical expectations}
The theoretical considerations to some extent follow the ideas developed
earlier for $B$-decays into 
pseudoscalar  (P) and vector (V) particles. We outline here two
complementary approaches.
\subsection{Phenomenological amplitudes}
The decay rates are expressed in terms of a set of phenomenological
amplitudes including the gluonic penguin, the electroweak tree
amplitudes and others. Such a scheme has been successfully applied to the decays 
$B\to PP,PV$ \cite{cr1,lipkin}. 

\begin{figure}[bht]
%\begin{center}
\vspace{-3.5cm}
\includegraphics[angle=+90,width=12cm,bbllx=240,bblly=88,bburx=427,bbury=749]{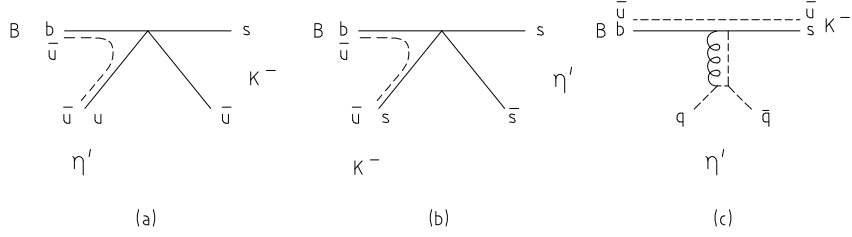}
%\includegraphics[angle=-90,width=14cm]{NF-dla.ps}
%\end{center}  
%
%\vspace{-5cm}
\caption{
Two-body decay $B^-\to K^-\eta'$ (or $B^-\to K^{*-}\eta'$) 
with three amplitudes from (\ref{penguin}):
 (a) amplitude $p^u_{K^-\eta'}$ with  $s\to K^-$, (b) exchange
amplitude $p^s_{\eta'K^-}$ with
 $s\to \eta'$ and (c) amplitude $s_{K^-\eta'}$ for
gluonic production of $\eta'$.}
\label{Ketaprime}
\end{figure}

Here we apply a scheme of this kind \cite{mo1}, 
but in this exploratory phase for scalars with
moderate statistics we restrict ourselves to including only the dominant
penguin diagrams and neglect in particular the tree diagrams which give rise
to corrections at the 20\% level. We then consider in this scheme 
the three $q\bar q$ processes with the
same amplitude as well as the gluonic amplitude
\begin{equation}
b\to s u\bar u, \quad b\to s d\bar d, \quad b\to s s\bar s , \quad b\to s g.
\label{penguin}
\end{equation} 
These processes together with the recombination of the spectator quark 
give rise to 2-body decays $B\to xy$ where $x,y$ are mesons out of the
flavour $U(3)$ nonets $A$ and $B$. Given the members of these multiplets
with a particular mixing angle the decay amplitudes can be given in terms
of the following parameters: the penguin amplitude $p_{AB}$ with $s\to x$, the exchange
amplitude $\beta_{AB}p_{AB}$ for $s\to y$ and $\gamma_{AB}p_{AB}$ for the
gluonic amplitude. For a more detailed discussion, see Ref. \cite{mo1}, here
we just adress for illustration the decay $ B^+\to \eta'K^{*+}$ for the
mixing $\eta'=(u\bar u, d\bar d, 2 s\bar s)/\sqrt{6}$. The decay amplitude
is derived from the penguin amplitudes as in Fig. \ref{Ketaprime} and reads
\begin{equation}
T=%\frac{1}{\sqrt{6}} p_{PV} + \frac{2}{\sqrt{6}}\beta_{PV} p_{PV}
%+\frac{4\gamma}{\sqrt{6}} p_{PV} \ = \
\frac{1}{\sqrt{6}}(1+2\beta_{PV}+4\gamma_{PV})p_{PV}
%\qquad \beta'=-\beta.
\label{etapr}
\end{equation}
A consequence of this penguin dominance model are various symmetry
relations, especially the $I=\frac{1}{2}$ rule: 
The final state of processes (\ref{penguin}) has $I=0$ and therefore the
final
state of $B$ decay has the isospin of the spectator, for $B^\pm,B^0$ 
this is $I=\frac{1}{2}$, which is also realized in our
amplitudes \cite{mo1}, see also Tab. \ref{tab:pseudosc} below.
%\vspace{-5cm}
%  
\subsubsection{Application to $B$-decays into pseudoscalars}
As a test of this penguin dominance model 
we have compared first with data on the decays $B\to
PP,\ B\to PV$ \cite{mo1}. In Tab. \ref{tab:pseudosc} this comparison is 
repeated with new data compiled by HFAG \cite{hfag}.
In col. 2 
we show our predictions 
for 12 decay rates of $B^+$ in terms of the parameters
$p_{PP},p_{VP}, \gamma_{PP},\gamma_{VP},\beta_{VP}$
and the corresponding 12 rates for $B^0$ obtained after multiplication
by the lifetime ratio $\tau(B^0)/\tau(B^+)=0.921$. From col. 2 various symmetry
relations can be obtained, especially the  
 $I=\frac{1}{2}$ rule (favouring charged $\pi$ or $\rho$
over the neutral decays by a factor 2)  for the doublets 
\begin{equation}
\begin{tabular}{llll}
$(K^0\pi^+),\ (K^+\pi^0)$;& $ (K^+\pi^-),\ (K^0\pi^0)$; &
 $(K^{*0}\pi^+),\ K^{*+}\pi^0)$;& $(K^{*+}\pi^-,K^{*0}\pi^0$)\\
$(\rho^+K^0,\rho^0K^+)$; & $(\rho^-K^+,\rho^0K^0)$ & & \\ 
\end{tabular}
%\label{isospinrule}
\nonumber
\end{equation}
These relations work well, except for one case
%$(K^{*+}\pi^-,K^{*0}\pi^0$) 
where the rate for $K^{*0}\pi^0$ is
significantly ($4.3\sigma$) below the expectation; however, the statistics
is very low in this case. Furthermore, there are SU(3) relations
between $K^*\pi$ and $\phi K$, and also between $\rho K$ and $\omega K$
which work reasonably well.

For a full description
we made some simplifying assumptions 
$\beta_{PV}=-1, \gamma_{PV}=\gamma_{PP}$ which can be removed if 
necessary with improving statistics. The remaining  3 parameters
$p_{PP},p_{VP},\gamma_{PP}$ have been determind from 3 input rates.
Remarkably, with the data of 
increased precision
obtained in the last year \cite{hfag} 
the agreement with the predictions has generally improved
in comparison to our earlier results in \cite{mo1} 
(2 exceptions with deviations of $> 3\sigma$). 

\begin{table}[t!]%[p]
\caption{
Branching ratios for $B$ decays into pseudoscalar (P) and
vector (V) particles (cols. 3,4,6,7) and amplitudes (col.~2) 
as in Eq. (\ref{etapr})  with 
$\gamma_{PP},\gamma_{VP}$ and $\beta_{VP}$ 
  for gluonic and interchange processes, 
$p_{PP}$, $p_{VP}$ set to 1;
cols.~3,6:  
$\gamma_{PP}=\gamma_{VP}=0.3$, $\beta_{VP}=-1$, $|p_{PP}|^2=24.1\times
10^{-6}$, $|p_{VP}|^2=11.4\times
10^{-6}$. Experimental Data from HFAG, July 2005 \cite{hfag}}
\vspace*{0.1cm}
$
\begin{array}{lclcllc}
\hline
B^+\to PP  & p_{PP}=1 & { B}_{\text{th}} [10^{-6}]
     &   { B}_{\text{exp}} [10^{-6}] & B^0\to PP & 
 { B}_{\text{th}} [10^{-6}]
     &   { B}_{\text{exp}} [10^{-6}]\\
\hline
%B^+ decays
K^0\pi^+ &  1 & \text{input}\ p_{PP} & 24.1\pm 1.3 & K^+\pi^- &  22.2 &
18.2\pm0.8  
\\
K^+\pi^0 & \frac{1}{\sqrt{2}} &  12.1 & 
     12.1\pm 0.8 & K^0\pi^0 & 11.1 & 11.5\pm 1.0\\
K^+\eta &   \frac{1}{\sqrt{3}}\gamma_{PP} 
     &  0.7 & 2.6\pm0.5 &K^0\eta & 0.7 & < 2.0
\\
K^+\eta' & \frac{1}{\sqrt{6}}(3+4\gamma_{PP})      &   \text{input}\ \gamma_{PP}
&70.8\pm3.4 &K^0\eta' & 65.3 & 68.6 \pm 4.2
\\

\hline
B^+\to VP & p_{VP}=1& & & B^0\to VP & & \\
\hline
% B+ decays
K^{*0}\pi^+ &   1  & \text{input}\ p_{VP} & 11.4\pm1.0 
  &K^{*+}\pi^- & 10.5& 11.7^{+1.5}_{-1.4}
\\
K^{*+}\pi^0 & \frac{1}{\sqrt{2}} & 
   5.7 & 6.9\pm2.3 & K^{*0}\pi^0 & 5.3 & 1.7\pm 0.8
\\
K^{*+}\eta & 
  \frac{1}{\sqrt{3}}(1-\beta_{VP}+\gamma_{VP})
      &  20.1 &  24.3^{+3.0}_{-2.9} &K^{*0}\eta & 18.5& 18.7\pm 1.7
\\
K^{*+}\eta' & 
\frac{1}{\sqrt{6}}(1+2\beta_{VP}+4\gamma_{VP}) &  0.8 & <14& K^{*0}\eta' &0.1& <7.6
\\
\rho^+ K^0 &  \beta_{VP} &  11.4 & <48  & \rho^- K^+ &10.5& 9.9^{+1.6}_{-1.5} 
\\
\rho^0 K^+ &  \frac{1}{\sqrt{2}}\beta_{VP}&
     5.7 &  5.11^{+0.82}_{-0.87} & \rho^0K^0 & 5.3 & 5.11^{+0.82}_{-0.87} 
\\
\omega K^+& \frac{1}{\sqrt{2}}\beta_{VP}&
       5.7 & 5.1\pm0.7  & \omega K^0 & 5.3& 5.6\pm 0.9
\\
\phi K^+ & 1 & 11.4& 9.03^{+0.65}_{-0.63} & \phi K^0 & 10.5&
8.3^{+1.2}_{-1.0}
\\
% B0 decays
%
\hline
\end{array}
$
\label{tab:pseudosc}
\end{table}

\subsubsection{$B$-decays into scalar particles}
After the success of this simple penguin dominance model 
we take it over to the decays with scalar particles
$B\to PS$ and $B\to VS$. We denote the members of the scalar multiplet 
by $a,\ K^*_{sc},\ f_0,\ f_0'$
and define the mixing angle by
%\begin{align}
%a&\ K^*_{sc},\ f_0,\ f_0'\\
$f_0=n\bar n \sin\varphi_s \ + \ s\bar s \cos\varphi_s$, 
$f_0'=n\bar n \cos\varphi_s \ - \ s\bar s \sin\varphi_s$,
%\end{align} 
where $n\bar n=(u\bar u+d\bar d)/\sqrt{2}$.
Then our predictions \cite{mo1} for scalars
are given in Tab. \ref{tab:scalars}. 
\begin{table}[thb]
%\parbox[h]{15.3cm}{
\caption{Dominant contributions for 
$B$ decays into scalar (S) + pseudoscalar (P) or
vector (V) particles: penguin amplitudes $p_{XY}$ (normalized to 1
in each sector),
exchange  and
gluonic amplitudes  $\beta_{PS},\beta_{VS}$ 
and $\gamma_{PS},\gamma_{SP},\gamma_{VS}$ 
resp. with scalar mixing angle $\varphi_S$; in brackets results for
 $\sin \varphi_S=1/\sqrt{3}\ (\varphi_S\sim\varphi_P)$;
%$\beta_{PS}=\pm1$,\  $\beta_{VS}=-\beta_{PS}$; 
cols. 3,6: upper sign for $B^0$,
lower sign $B^+$.}
%; notations
%$f_0\equiv f_0(980),\ f_0'\equiv f_0(1500)$ and $K^*_{sc}\equiv K^*_0(1430)$.}
%\vspace*{0.1cm}
$
\begin{array}{llcllc}
% \begin{array}{ll@{\hspace*{3.2cm}}c@{\hspace*{3.2cm}}c}
\hline
B^0\to & B^+\to & \text{normalization to}&
  B^0\to & B^+\to & \text{normalization to}\\
P+S & P+S &  p_{PS}=1 & V+S & V+S &  p_{VS}=1 \\
 \hline
K^+a^- & K^0 a^+ & 1 & K^{*+}a^- & K^{*0} a^+ & 1 
\\
K^0a^0 & K^+ a^0 & \mp\frac{1}{\sqrt{2}} & K^{*0}a^0 & K^{*+} a^0 &
   \mp\frac{1}{\sqrt{2}} 
\\
K^0f_0 & K^+f_0 & \frac{1}{\sqrt{2}}(1+2\gamma_{PS})\sin \varphi_S &
  K^{*0}f_0 & K^{*+}f_0 & \frac{1}{\sqrt{2}}(1+2\gamma_{VS})\sin \varphi_S\\
       &        & \quad +(\beta_{PS} + \gamma_{PS})\cos \varphi_S &
       &        & \quad  +(\beta_{VS} + \gamma_{VS})\cos \varphi_S  \\
%     & & \approx  \frac{1}{\sqrt{6}}(3+4\gamma_{PS}) & 
     & & \lbrack \frac{1}{\sqrt{6}}(1+2\beta_{PS}+4\gamma_{PS}) \rbrack &
%     & & \approx  \frac{1}{\sqrt{6}}(-1+4\gamma_{VS})\\
     & & \lbrack \frac{1}{\sqrt{6}}(1+2\beta_{VS}+4\gamma_{VS})\rbrack \\
K^0f_0' & K^+f_0' & \frac{1}{\sqrt{2}}(1+2\gamma_{PS})\cos \varphi_S & 
     K^{*0}f_0' & K^{*+}f_0' & \frac{1}{\sqrt{2}} (1+2\gamma_{VS})\cos
      \varphi_S\\
      & & \quad   -(\beta_{PS} + \gamma_{PS})\sin \varphi_S &
      & & \quad  -(\beta_{VS} + \gamma_{VS})\sin \varphi_S  \\
%     & & \approx  \frac{1}{\sqrt{3}} \gamma_{PS} &
     & &  \lbrack \frac{1}{\sqrt{3}}(1-\beta_{PS}+\gamma_{PS}) \rbrack &
%         & & \approx  \frac{1}{\sqrt{3}} (2+\gamma_{VS})\\
        & & \lbrack \frac{1}{\sqrt{3}} ( 1-\beta_{VS}+\gamma_{VS})\rbrack \\
\pi^-K^{*+}_{sc} & \pi^+K^{*0}_{sc} & \beta_{PS} & 
    \rho^- K^{*+}_{sc} &  \rho^+ K^{*0}_{sc} & \beta_{VS}
\\
   \pi^0K^{*0}_{sc} & \pi^0K^{*+}_{sc} & \mp\frac{1}{\sqrt{2}} \beta_{PS} &
   \rho^0K^{*0}_{sc} & \rho^0K^{*+}_{sc} & \mp\frac{1}{\sqrt{2}} \beta_{VS}
\\
\eta K^{*0}_{sc} & \eta K^{*+}_{sc} &
\frac{1}{\sqrt{3}}(-1+\beta_{PS}+\gamma_{SP})&
     \omega K^{*0}_{sc} & \omega K^{*+}_{sc} & \frac{1}{\sqrt{2}} \beta_{VS} 
\\
\eta' K^{*0}_{sc} & \eta' K^{*+}_{sc} &
    \frac{1}{\sqrt{6}}(2+\beta_{PS}+4\gamma_{SP})&
   \phi  K^{*0}_{sc} & \phi K^{*+}_{sc} & 1\\
\hline
  \end{array}
$
\label{tab:scalars}
%}
\end{table}
Given the decay branching ratios into scalars one can check any scenario for
the multiplet of scalar particles. Hopefully, the symmetries implied by
penguin dominance (isospin, SU(3)) inherent in Tab. \ref{tab:scalars} 
will help in selecting the correct assignments of scalar particles. 
The parameters we have at our disposal are for $B\to PS$:
$p_{PS},\gamma_{PS},\gamma_{SP},\beta_{PS}$ and for $B\to VS$: 
$p_{VS},\gamma_{VS},\beta_{VS}$. In our first analysis \cite{mo1}
we used initially, in analogy to the pseudoscalars, $\beta_{PS}=1$, 
$\beta_{VS}=-\beta_{PS}$, $\gamma_{VS}=\gamma_{PS}=\gamma_{SP}$.

\subsubsection{Comparison with experimental results on scalars in $B$ decays}

Considering first the multiplet 
 $\sigma,\kappa,f_0(980),\ a_0(980)$ along scenario I we note that only
$f_0(980)$ has been observed so far.
For a meaningful test one would need a measurement of the rates 
for $B\to K\sigma$ and $B\to \pi \kappa$ which should be possible for a
given parametrization. 

On the other hand, the decay rates for all members of the multiplet along
scenario II  $f_0(980),\ a_0(980),\ K^*_0(1430),\ f_0(1500)$ 
have actually been measured (upper limit
for $a_0$ only).
According to our scheme with penguin dominance we should describe these four
rates by 3 parameters: $p_{PS},\gamma_{PS}$ and $\beta_{PS}$.

In a first attempt in 2004 we analysed the data assuming as in case of
pseudoscalars $|\beta_{PS}|=1$. Then we expected for the decay 
$B(B\to K a_0^+)\gsim
11.0\times 10^{-6}\ (\pm 50\%)$ The new upper limits from BaBar in 
Tab. \ref{babara0dat} are below this expectation. From Tab. \ref{tab:scalars} we
find $B(B\to K^*_0\pi^\pm)/B(B\to Ka^\pm)=|\beta_{PS}|^2$. 
The new data require
$|\beta_{PS}|\sim 2.7\ldots 4.6$, or, from averages using the $I=1/2$ rule
$|\beta_{PS}|\gsim 2$. The production of a scalar with the spectator is
suppressed against production from $s$-quark. 

Until now, 
there are still considerable experimental uncertainties, 
especially the ambiguities in the $K^*_0(1430)$ rates and the
missing $a_0$ rate. If we choose $|\beta_{PS}|=2$ then we find with $B(B\to
K^*_0\pi^+)\sim 12\times 10^{-6}$ (if we include the lower $K^*_0$ mass), 
for $B(B\to a_0^+K^0)\sim 3 \times 10^{-6}$
and $B(B\to Kf_0(980))\sim 13 \times 10^{-6}$ four solutions in ($\beta,\gamma$). For
$\beta=-2,\gamma=2$ we find $B(B\to Kf_0(1500))\sim 25\times 10^{-6}$ which
compares well with Sol. II in Tab. \ref{bf01500}. So there is no difficulty in the
moment with the multiplet along path II considered. The tests will hopefully
become more restrictive with improved data and with measurements of other
channels like $K^*_0(1430)\eta$,  $K^*_0(1430)\eta'$ and $B\to V+$ scalars.

\subsection{QCD-improved factorization approximation}
In this complementary theoretical investigation one aims at an absolute
prediction of rates for scalar particles \cite{chernyak,furman,cheng}.
This follows the approach applied before to decays $B\to
PP,\ VP$ \cite{bbns}. In the recent work \cite{cheng} 
one includes perturbative QCD corrections to the common
factorization ansatz but needs to include various non-perturbative objects:
formfactors, light cone distribution amplitudes and decay constants where
results for scalars are derived from QCD sum rules. In scenario I 
$\sigma, \kappa,a_0(980),f_0(980)$ are taken as $q\bar q$ ground states and
$a_0(1450), K^*(1430), f_0(1500) $ as  $q\bar q$ excited states. In scenario
II it is assumed that the low mass multiplet is build of $qq\bar q\bar q$
states for which no quantitative predictions can be given, whereas the $q\bar
q$ ground state multiplet includes $a_0(1450), K^*_0(1430)$ and a second
multiplet is around 2 GeV. 

An early calculation \cite{chernyak} predicted a very small rate for $B^0 \to
a_0^+K^- $ which turned out successful. The recent predictions \cite{cheng}
concern  decays into $a_0(980)$, $f_0(980)$, also $K^*(1430)$ and
$a_0(1430)$. Within scenario I the results on the low mass multiplet are
satisfactory whereas the higher mass particles require the low mass
solutions with $B(B^-\to K^*(1430)^-\pi^+) < 10 \times 10^{-6}$.
In scenario II the $K^*_0$ rates are about twice as large as before, but
still smaller than some experimental results. 

If this large $K^*_0$ rate is
correct, then scenario I is excluded and there are no predictions for 
the light mesons with $M\lsim 1$ GeV.
It will be important to know the predictions for the other
states $\sigma,\kappa$ to compare with, likewise predictions for
$f_0(1500)$ and the other isoscalar meson.

\section{Conclusions}

\emph{1. Experimental results on decays $B\to $ Scalar $ + X$}\\
By now  $f_0(980)$, $a_0(980)$,
 $K^*_0(1430)$ and $f_0(1500)$ have been measured in $B$-decays. 
Discrete ambiguities
are found for $K^*_0(1430)$, $f_0(1500)$ (how about
$f_0(980)$?) and emerge naturally in coherent superposions. 
A clarification is important, possibly these ambiguities can be 
resolved by physical arguments
(comparison with 
elastic scattering phases, isospin relations fulfilled within $\sim 20$\%). 

\emph{2. Model with gluonic penguins dominating and $B\to S+X$ amplitudes}\\
This model continues to work well for $B\to PP,VP$ 
within $\sim 20\%$ or better, especially the $I=\frac{1}{2}$ rule 
and other SU(3) relations are generally successful. The method
has the potential to test the multiplet structure in the scalar sector.
Present data within their ambiguities are consistent 
with a $q\bar q$ multiplet
$f_0(980), a_0(980), K^*_0(1430), f_0(1500)$.
Further tests are possible with $B\to K^*_0(1430)\eta$ (or $\eta'$) as well as  
$VS$ rates. The possibility of a light multiplet with  $\sigma,\kappa$
can be tested once data on $B\to \sigma K,\kappa \pi$ become available. 

\emph{3. Factorization approach for $B$-decays into scalar particles}\\ 
Using QCD sum rules to obtain nonperturbative quantities 
some absolute predictions have been obtained, a successful one concerns the
decay into $a_0(980)$. Further distinctions between different scenarios
depend on the magnitude of the ambiguous $K^*_0(1430)\pi$ rate.   
It will be important to have predictions for the other members 
of the considered
multiplets, especially for $B\to \sigma K, \kappa\pi$, as well as for heavier
isoscalars.

\emph{4. Broad state: a respectable 
glueball candidate and the $X(1500), f_X(1500)$
puzzle.}\\
In the $\pi\pi$ channel there is a broad state with $\Gamma\sim M$. It is
plausible that $\sigma(600)$ and $f_0(1000)$ refer to the same object.
The puzzles with $X(1500), f_X(1500)$ are resolved by taking into
account the interference of $f_0(1500)$ with a 
broad background. The relative signs are explained by taking the background
 as flavour singlet, in agreement with the glueball hypothesis,
and $f_0(1500)$ as a flavour octet state. The same interference phenomenon is
known from processes $\pi\pi\to \pi\pi,\eta\eta, K\bar K$.

\begin{theacknowledgments}
Thanks to the organizers  for creating a
very lively conference with ample opportunities to in depth discussions!
\end{theacknowledgments}

\section{Note added}
After this conference a paper by Gronau and Rosner \cite{gronau} 
appeared with isospin relations  between pairs of $B^0$ and $B^+$ 
2-body decays as well as 3-body decays also basing on the dominance 
of penguin amplitudes.



%\bibliography{sample}

\IfFileExists{\jobname.bbl}{}
 {\typeout{}
  \typeout{******************************************}
  \typeout{** Please run "bibtex \jobname" to optain}
  \typeout{** the bibliography and then re-run LaTeX}
  \typeout{** twice to fix the references!}
  \typeout{******************************************}
  \typeout{}
 }

%%%%%%%%%%%%%%%%%%%%%%%%%%%%%%%%%%%%%%%%%%%
%% The following lines show an example how to produce a bibliography
%% without the help of the BibTeX program. This could be used instead
%% of the above.
%%%%%%%%%%%%%%%%%%%%%%%%%%%%%%%%%%%%%%%%%%%

\end{document}